\documentclass[conference]{IEEEtran}
\IEEEoverridecommandlockouts

\usepackage{amsmath,amssymb,amsfonts}
\usepackage{algorithmic}
\usepackage{graphicx}
\usepackage{textcomp}
\usepackage{xcolor}
\usepackage{booktabs}
\usepackage{cleveref}
\usepackage{amsmath}
\usepackage{algorithm}

\def\BibTeX{{\rm B\kern-.05em{\sc i\kern-.025em b}\kern-.08em
    T\kern-.1667em\lower.7ex\hbox{E}\kern-.125emX}}

\usepackage[style=ieee, backend=biber, mincitenames=1, maxcitenames=1]{biblatex}
\begin{document}

\title{
EquivAnIA:
A Spectral Method for Rotation-Equivariant Anisotropic Image Analysis
}

\author{
    \IEEEauthorblockN{
        \begin{tabular}{cc} 
            Jérémy Scanvic${}^{\dagger, 1,2}$ & Nils Laurent${}^{\dagger,3}$
            \thanks{${}^\dagger$ The two authors have contributed equally to this work.}
        \end{tabular}
    }
    \IEEEauthorblockA{${}^1$ Laboratoire de Physique de l'ENS de Lyon, CNRS, Lyon, France ${}^2$ Hirsch, Lyon, France \\ ${}^3$ Université Jean Monnet, LASPI, Roanne, France}
}

\maketitle

\begin{abstract}
Anisotropic image analysis is ubiquitous in medical and scientific imaging, and while the literature on the subject is extensive,
the robustness to numerical rotations of numerous methods remains to be studied.
Indeed, the principal directions and angular profile of a rotated image are often expected to rotate accordingly.
In this work, we propose a new spectral method for the anisotropic analysis of images (EquivAnIA) using two established directional filters, namely cake wavelets, and ridge filters.
We show that it is robust to numerical rotations throughout extensive experiments on synthetic and real-world images containing geometric structures or textures, and we also apply it successfully for a task of angular image registration\footnote{The code is available at \url{https://github.com/jscanvic/Anisotropic-Analysis}}.
\end{abstract}


\section{Introduction}


Anisotropic image analysis is ubiquitous in medical and scientific imaging~\cite{lafarge20RotoTranslation}, yet the design of methods with a satisfying degree of robustness to rotations remains an active research subject~\cite{pu23Adaptive}.

One of the foundational tools from spectral analysis is the two-dimensional power spectral density (PSD)~\cite{vetterli14Foundations},
in which the power is spread across each pair of vertical and horizontal frequency.
By integrating it over the radius in a polar coordinate system, it yields the angular PSD, which encodes the different directions of anisotropy of the input image~\cite{hu18Fourier}.

However, in practice, the power spectral density is generally estimated on a Cartesian grid in practice,
and estimating the angular PSD with robustness to rotations remains a challenging task.
Indeed, there are many ways to integrate the discretized PSD over arbitrarily oriented lines.

In this work, we seek to express the anisotropy from the two-dimensional PSD of an image, which we call the angular profile.
In particular, we consider three different approaches using cake wavelets~\cite{bekkers14MultiOrientation}, ridge filters, and angular binning.

This work specifically focuses on single-resolution anisotropic image analysis, and we leave to future work the evaluation of the empirical robustness to rotations of tools from multi-resolution anisotropic analysis, including ridgelets~\cite{donoho01Ridge}, curvelets~\cite{candes00Curvelets} and shearlets~\cite{guo06Sparse}.



We apply our findings to the task of angular image registration, which consists in estimating the rotation angle between two rotated copies of the same base image.


Our contributions are the following:
\begin{itemize}
    \item We propose a new spectral method for numerical anisotropic image analysis.
    \item We validate our method and demonstrate its robustness to numerical rotations on synthetic and real-world images containing geometrical structures and textures.
    \item We evaluate it against competing methods on a task of angular image registration.
\end{itemize}








\section{Background} \label{sec:background}

The power spectral density of an image $x(u)$, denoted as $S(\xi)$, is a powerful image descriptor~\cite{vetterli14Foundations}.
It is generally estimated as the image periodogram
\begin{equation} \label{eq:periodogram}
    S(\xi) = S(\xi_1, \xi_2) = |\widehat x(\xi)|^2,
\end{equation}
where $\widehat x(\xi)$ denotes the Fourier transform of $x(u)$, and $\xi_1$ and $\xi_2$ denote respectively horizontal and vertical frequencies.

For images that have non-periodic boundary conditions, a windowing step is generally performed before computing the discrete Fourier transform (DFT) to remove ringing artifacts.
In certain cases, periodograms computed on possibly overlapping patches are averaged to obtain a PSD with reduced noise, at the expense of a loss in resolution, this is notably the case in Bartlett's and Welch's methods.

Even though it is generally defined in a Cartesian coordinate system, the PSD can also be expressed in polar coordinates as
\begin{equation}
    S(r, \theta) = r \cdot S(r \cos \theta, r \sin \theta),
\end{equation}
for anisotropic image analysis applications~\cite{abbas17Exact}.
By integrating it along the radius coordinate, we obtain the angular power spectral density (angular PSD)
\begin{equation}
    S(\theta) = \int_0^\infty S(r, \theta) \mathrm d r,
\end{equation}
which associates to each angle part of the total image power.

In the discrete setting, it is approximated by first associating each pair of horizontal and vertical frequency to an angular bin based on its angle in polar coordinates, then summing the PSD values in each bins to replicate the integration step.
In this work, it is referred to as the binning method as each frequency is used to compute the power in a single direction, thereby preserving the total power.

\begin{figure}[th]
    \centering
    \includegraphics[width=0.75\linewidth]{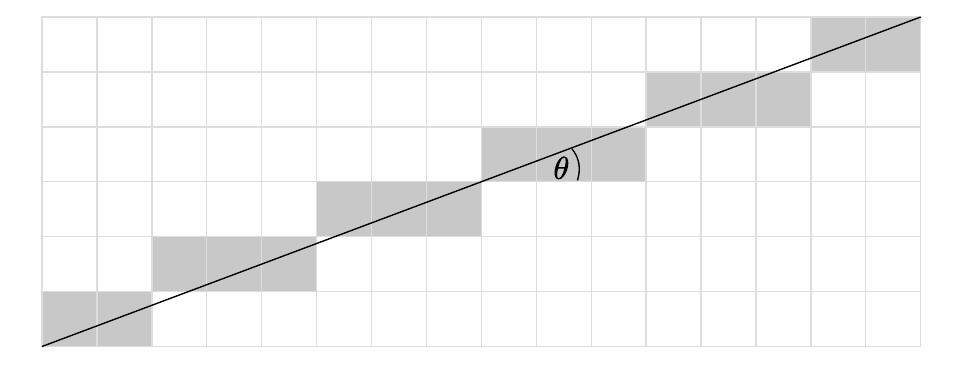}
    \caption{Illustration of selected pixels using the binary line rasterization obtained with Bresenham algorithm.}
    \label{fig:raster_line}
\end{figure}

Equivalently, it can be understood as the inner product of the discrete PSD with a weighted mask defined in the Fourier domain, consisting in the rasterization of a line starting at the zero frequency~\cite{tunak07Analysis}.
\Cref{fig:raster_line} illustrates the possible rasterization of a line, using Bresenham algorithm.

The angular bins have varying sizes due to the anisotropic grid the DFT is sampled on, e.g., the 0\textdegree angle is associated with more frequencies than the 30\textdegree angle.
Hence, the binning method typically results in different angular profiles for different rotations of the same input, which is not desirable in many applications.

In \Cref{sec:method}, we overcome this limitation and propose a method for anisotropic image analysis that is robust to input rotations.








\section{Anisotropic image analysis} \label{sec:method}


In this section, we introduce our proposed method for anisotropic image analysis, which makes use of the same spectral tools as the binning method described in~\Cref{sec:background}. It consists in three successive steps that we present hereafter.

Instead of estimating a discretization of the angular PSD $S(\theta)$ using angular binning, we consider an angular profile $\rho(\theta)$ of the input image defined, for each angle $\theta \in [0, \pi)$, as a weighted average of its PSD values, for each angle, in the vicinity of the angle $\theta$.

We find empirically that Bartlett's and Welch's methods for PSD estimation lead to poorer anisotropic analyses, and we believe that it is due to the loss of resolution they incur. Hence, we use the periodogram estimation of the PSD, which is defined in~\cref{eq:periodogram}.


Except for images supported on a disk, like tomography scans, we apply a smooth radially-symmetric window approximately supported on a disk before the PSD is computed.
It makes the estimation of the PSD more robust to rotations as it discards information that may enter or leave the corners of the image domain as the input rotates.





\begin{figure}[thbp]
    \centering
    \includegraphics{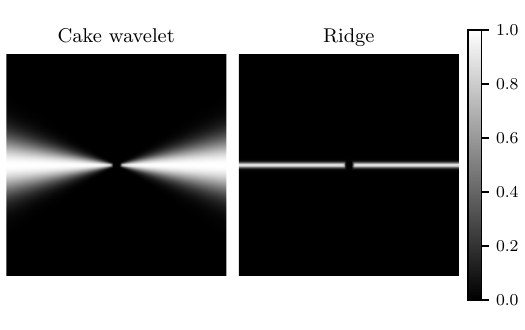}
    \caption{\textbf{Filter spectra.} The Fourier transform of the filters functions used in our method.}
    \label{fig:filters}
\end{figure}

Analysis of input images $x(u)$ is done using a family of oriented functions $\phi_{v,\theta}(u)$
where $v \in \mathbb R^2$ denotes a position in the image plane and $\theta \in [0, \pi)$ an orientation.
We generate them from a base function $\phi := \phi_{v_0, \theta_0}$ where $v_0 = 0$ and $\theta_0 = 0$, transformed by translations and rotations
\begin{equation}
    \phi_{v, \theta}(u) = \phi\left(R_\theta^{-1} (u - v)\right), \quad u \in \mathbb R^2,
\end{equation}
where $R_\theta \in \mathbb R^2$ is the rotation matrix of angle $\theta$. The analysis coefficients associated to this family of functions is denoted as
\begin{equation}
    c_{v,\theta} = \int_{\mathbb R^2} x(u) \bar{\phi}_{v, \theta}(u) \mathrm du, \ v \in \mathbb R^2,\ \theta \in [0, \pi),
\end{equation}
where $\bar{\phi}_{v, \theta}(u)$ is the complex conjugate of $\phi_{v, \theta}(u)$.
The angular profile is defined as the energy response at each orientation,
\begin{equation}
    \rho(\theta) = \int_{\mathbb R^2} |c_{v,\theta}|^2 \mathrm d v, \quad \theta \in [0, \pi).
\end{equation}

In practice, the base analysis function is either a cake wavelet filter or a ridge filter parametrized directly in the Fourier domain\footnote{Their multi-resolution analogues are curvelets~\cite{bekkers14MultiOrientation} and ridgelets~\cite{donoho01Ridge}.}, which are shown on~\Cref{fig:filters}. Note that the filters are rendered centrally-symmetric in the Fourier domain as we weight angles that are 180\textdegree{} apart the same.



Anisotropic images are likely to exhibit a principal orientation $\eta \in [0, \pi)$, i.e., a direction along which its power is greater, and we estimate it from the angular profile $\rho(\theta)$ as
\begin{equation} \label{eq:orientation_estimate}
    \eta = \mathop{\mathrm{argmax}}_{\theta \in [0, \pi)} \ \rho(\theta).
\end{equation}
Even though the estimated angular profile may theoretically have multiple global maxima, we find that it is empirically not the case for images with a clear principal direction of anisotropy.


\section{Angular registration} \label{sec:registration}

The problem of angular registration consists in identifying the relative rotation between two rotated copies of the same image $x^{(1)}$ and $x^{(2)}$, i.e., we are looking for $\gamma \in [0, 2\pi)$ such that
\begin{equation}
    x^{(2)} = R_\gamma x^{(1)}.
\end{equation}

Existing approaches for registration include SIFT registration~\cite{lowe99Object}, \cite{reddy1996fft}, and spectral and Radon methods~\cite{albu16Low,reddy1996fft}.

\begin{figure*}[thbp]
    \centering
    \includegraphics[width=.97\textwidth]{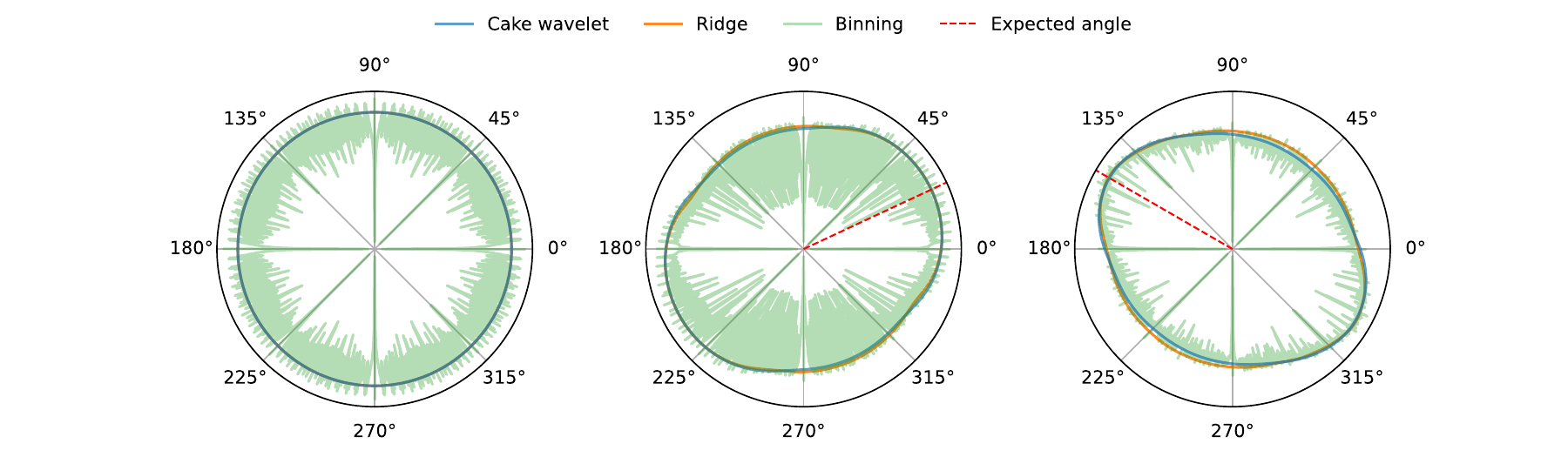}
    \caption{
    \textbf{Angular profiles.} For each image in~\Cref{fig:illustration_img},  the predicted profiles using our method and the baselines are displayed.
    }
    \label{fig:illustration_analysis}
\end{figure*}

Our method consists in first computing the orientation estimates defined in~\cref{eq:orientation_estimate} for each of the two images, denoted as $\widehat \theta^{(1)}$ and $\widehat \theta^{(2)}$ respectively. Similarly to~\textcite{keller05Angular}, in order to overcome the fact that our angular estimation does not distinguish an image and its 180\textdegree{} rotation, we test two possible registration angles $\widehat \gamma_1$ and $\widehat \gamma_2$ defined as
\begin{equation}
    \widehat \gamma_1 = \widehat \theta^{(1)} - \widehat \theta^{(2)}, \quad \widehat \gamma_2 = \widehat \theta^{(1)} - \widehat \theta^{(2)} + \pi \mod 2\pi.
\end{equation}
the final registration estimate $\widehat \gamma \in [0, 2\pi)$ is determined by that which give the best mean squared error
\begin{equation}
    \widehat \gamma \in \mathop{\mathrm{Argmin}}_{k \in \{ 1, 2 \}} \quad \left\| x^{(1)} - R_{\gamma_k} x^{(2)} \right\|^2.
\end{equation}

Despite being unable to distinguish between angles $\theta$ and $\theta + \text{180\textdegree}$, our method remains powerful as demonstrated by our registration experiments. Moreover, it could be extended for use with standard tools like the Hilbert transform to gain the lost granularity.


The end-to-end method is summarized in~\Cref{alg:registration_algorithm}.


%
%


\section{
Experiments
} \label{sec:experiments}

In this section, we present our experiments on anisotropic image analysis using synthetic images with known anisotropic properties, and real-world images.

\paragraph{Methods}
We compare two variants of our method with cake wavelets and ridge filters, described in~\Cref{sec:method}, to the binning baseline described in~\Cref{sec:background}.

\begin{figure}[hbtp]
    \centering
    \includegraphics[width=0.15\textwidth]{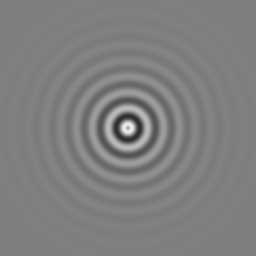}
    \includegraphics[width=0.15\textwidth]{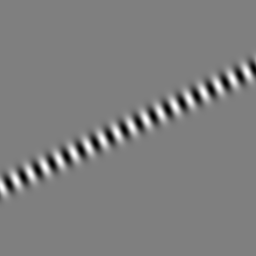}
    \includegraphics[width=0.15\textwidth]{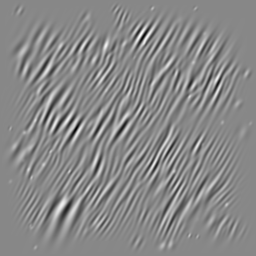}\\
    \vphantom{.} \hfill (a) \hspace{2.25cm} (b) \hspace{2.25cm} (c) \hfill \vphantom{.}\\
    \vspace{0.2cm}
    \includegraphics[width=0.15\textwidth]{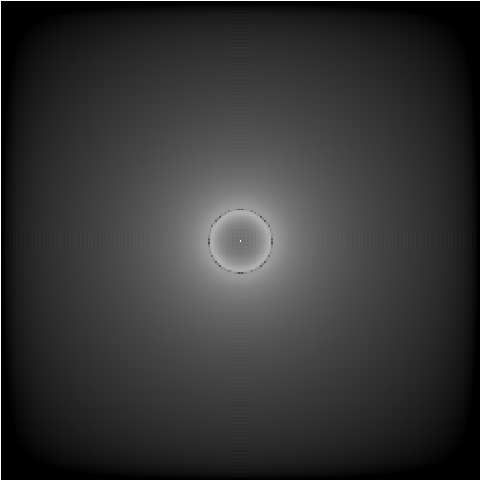}
    \includegraphics[width=0.15\textwidth]{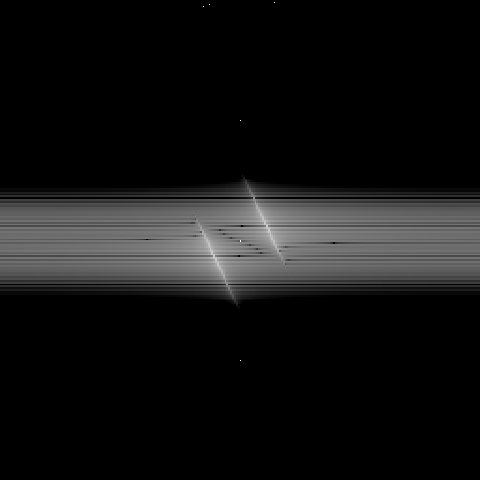}
    \includegraphics[width=0.15\textwidth]{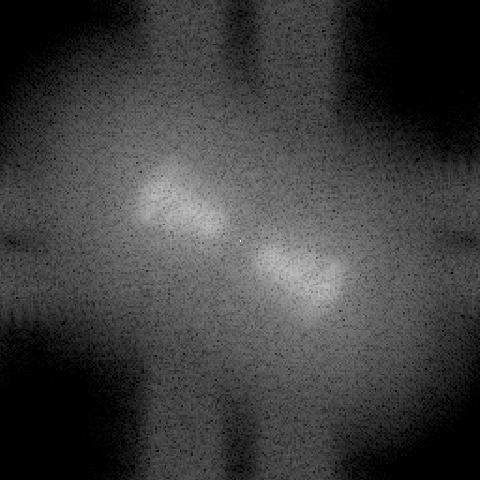}\\
    \vphantom{.} \hfill (d) \hspace{2.25cm} (e) \hspace{2.25cm} (f) \hfill \vphantom{.}
    \caption{
    \textbf{Example of synthetic images used in the experiments.}
    The spectrum in logarithmic scale of each image is shown below it.
    }
    \label{fig:illustration_img}
\end{figure}

\paragraph{Metrics}
We compare the different methods using performance metrics and metrics of rotation-equivariance to assess their robustness to changes in orientation.
In the experiments, when synthetic images are anisotropic, we compute a distance to the true angular profile, which can be the mode of a von-Mises distribution when using random Gabor atoms, or an angle parameter in the case of~\Cref{fig:illustration_img} (b).

We also compare the angular distance in degrees between the true principal orientation and the estimated principal orientation.
The distance used to compare angular profiles is the mean squared error expressed in dB.
In the experiments on image registration, we compute the angular distance between the true and estimated registration angles, in degrees, and the average equivariance error in degrees when applying a random rotation to the two input images.

\begin{algorithm}
\caption{Registration algorithm}
\label{alg:registration_algorithm}
\begin{algorithmic}[1]
\REQUIRE Input images $x^{(1)}, x^{(2)}$, window $w$

\FOR{$k \in \{ 1, 2 \}$}

\STATE $\widehat{x}^{(k)} \gets \text{FFT2}\left(w \odot x^{(k)}\right)$

\STATE $S^{(k)} \gets \left|\widehat{x}^{(k)}\right|^2$

\FOR{$\theta = \theta_1, \dots, \theta_M$}
    \STATE $\rho^{(k)}(\theta) \gets \sum_{l=1}^N \sqrt{\phi_{l, \theta}} \cdot S_l^{(k)}$
\ENDFOR

\STATE $\widehat{\theta}^{(k)} \gets \mathop{\mathrm{argmax}}_{\theta} \rho^{(k)}(\theta)$

\ENDFOR

\RETURN $\widehat \gamma \gets \mathop{\mathrm{argmin}}_{\delta \in \{ 0, \pi \}} \left\| x^{(1)} - R_{\widehat \theta^{(1)} - \widehat \theta^{(2)} + \delta} x^{(2)} \right\|^2$
\end{algorithmic}
\end{algorithm}

\begin{table}[hbtp]
    \centering
    \caption{\textbf{Performance on synthetic images.} The angular distance is in degrees, and the profile distance is a mean squared error in dB.}
    \setlength{\tabcolsep}{11pt} 
    \begin{tabular}{lcc}
        \toprule
         Method & Angular distance $\downarrow$ & Profile distance $\uparrow$ \\
         \midrule
        Cake wavelet & \textbf{0.03 ± 0.25} & \textbf{94.47 ± 2.50} \\
        Ridge & 0.06 ± 0.35 & 88.08 ± 2.26 \\
        Binning & 0.32 ± 0.84 & 50.79 ± 1.08 \\
         \bottomrule
    \end{tabular}
    \label{tab:statistical_analysis}
\end{table}

\paragraph{Synthetic images}

In this section, we illustrate the performance of each method on three synthetic images. Each image and its spectrum in provided in first row and second row of~\Cref{fig:illustration_img} respectively. The predicted angular profiles of each images are given in~\Cref{fig:illustration_analysis} left, center and right, where each profile is normalized by its sum.

\begin{figure}[bhtp]
    \centering
    \includegraphics[width=0.24\textwidth]{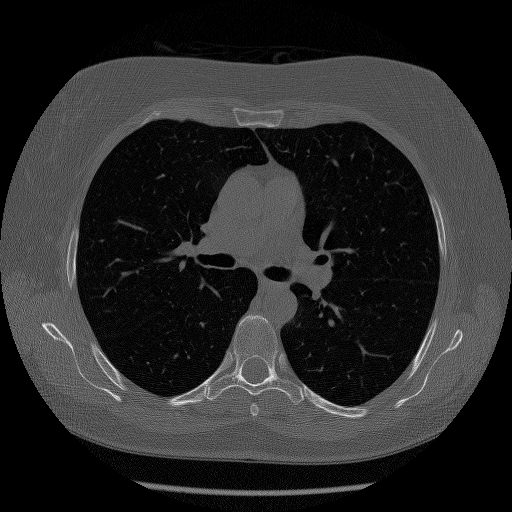}
    \includegraphics[width=0.24\textwidth]{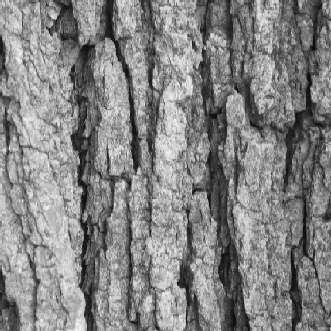}
    \caption{\textbf{Real-world images used in the experiments.} (left) a CT scan from the LIDC-IDRI dataset~\cite{armatoiii11Lung}, (right) the bark of a tree.}
    \label{fig:real_img}
\end{figure}

\begin{figure}[hbtp]
    \centering
    \includegraphics[width=1.00\columnwidth]{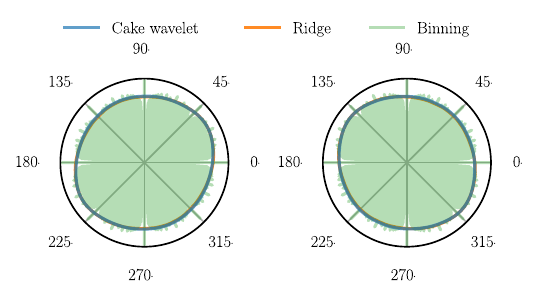}
    \caption{
    \textbf{Angular profiles in image registration.} (Left) First image, (right) Second image, both obtained by rotating the bark texture.
    }
    \label{fig:profiles_registration}
\end{figure}

In the case of~\Cref{fig:illustration_img} (a), the image is isotropic and since its this kind of images have an isotropic spectrum, we expect a constant angular PSD. In~\Cref{fig:illustration_analysis} (left), we observe that the best profiles are given by the Cake wavelet and the Ridge methods, which are the most constant whereas the binning method varies much more.

In the case of~\Cref{fig:illustration_img} (b), the image contains oscillations along a line with an angle of $25$°, and in its spectrum, the power is also located in that same direction. In~\Cref{fig:illustration_analysis} (center), we see that the smoothest profiles are again given by the Cake wavelet and Ridge methods, but this time we also remark that their maximum value is taken in the neighborhood of $\theta = 25$°. In the case of~\Cref{fig:illustration_img} (c), the image is a linear combination of $L = 300$ Gabor atoms. Each atom depends on the following function
\begin{equation}
    h_{\theta}(a, b) = \exp \left(-a^2 - b^2\right) \cos \left(a \sin \theta - b \cos \theta \right),
\end{equation}
which oscillates locally around zero at angle $\theta$. Then, each Gabor atom can be expressed as
\begin{equation}
    g_{\theta, s, u, v} = h_{\theta} \left(\frac{a - u}{s}, \frac{b - v}{s}\right),
\end{equation}
where the parameters $s, u, v$ are uniformly distributed on finite intervals. The parameter $\theta$ follows a von-Mises distrubtion shifted by $\pi$ and scaled by $\frac12$ to obtain an angle in $[0, \pi)$. This distribution has two parameters : a center $\mu$ and a concentration $\sigma$.
Image of \Cref{fig:illustration_img} (c) was generated with $\mu = \frac{\pi}{3} = 60$°, and we observe that its spectrum contains power in a cone centered on $\mu$. Here again, the profiles in \Cref{fig:illustration_analysis} (right) show that the cake wavelet and Ridge methods are smoothest and take their maximum value around $\mu$.

We propose a statistical study of the methods from the results in \Cref{tab:statistical_analysis}. The statistics were computed from $N = 300$ synthetic images obtained by summing $L = 300$ gabor atoms, i.e. each image is distributed similarly to \Cref{fig:illustration_analysis} (c). In this table, the methods are compared from their average and standard deviation on two metrics: the angular distance between the principal estimated angle and the parameter $\mu$, and the profile equivariance error.

We observe that for both metrics, the cake wavelet method has the lowest distance and the lowest variance and is therefore the best one. In addition, we observe that the ridge method gives acceptable results relatively close to the cake wavelet method. The Binning method performs much worse on all metrics, since it has significantly higher values of distances and variances.

\paragraph{Real-world images}

In addition to the experiments on synthetic images, where ground truth angular profiles are perfectly known, we evaluate our method on two real-world images with unknown ground truth angular profiles.
Namely, we use a CT scan from the LIDC-IDRI dataset~\cite{armatoiii11Lung}, and the photograph of the bark of a tree, which covers both structure and texture images. \Cref{fig:real_img} shows the two images.


We apply two bilinear rotations with arbitrary angles to each two images to obtain two pairs of images differing only by their orientation.
In the noisy setting, we further corrupt the two resulting images with additive white Gaussian noise with a SNR of 25~dB.

We then apply the algorithm in~\Cref{sec:registration} with the different profile estimation methods.
\Cref{tab:registration} shows that our method outperforms the binning method which is unable to provide faithful angular profiles due to its lack of robustness to numerical rotations.
Moreover, the cake wavelet variant of our method appears to perform better on structure images, while the ridge variant of our method appears to perform better on texture images.

\Cref{fig:profiles_registration} shows the different estimated profiles for the bark image, and we observe that while the two variants of our method are smooth and rotate with the input rotation, the binning baseline does not and is biased towards the grid-aligned angles like the 0\textdegree, 45\textdegree, and 90\textdegree angles.

Additionally, \Cref{tab:robustness} shows that our method remains performant even when each image is corrupted with additive white Gaussian noise, and therefore applies even when the two images are not exact rotations of a base image. This is expected as the noise has the same power in every direction and thus it does not introduce any bias in the angular estimation. We leave the more challenging study of robustness to non-isotropic noise to future work.




\begin{table}[thbp]
    \centering
    \caption{\textbf{Registration performance on real-world images.}
    The registration and equivariance error are expressed in degrees.
    The best metric is in \textbf{bold}.
    }
    \label{tab:registration}
    \setlength{\tabcolsep}{7pt} 
    \begin{tabular}{llcccccc}
    \toprule
    Image & Method & Reg. err. $\downarrow$ & Equiv. err. $\downarrow$ \\
	\midrule
	  CT scan & Cake wavelet & \textbf{0.02} & 0.47 \\
	  & Ridge & 0.16 & \textbf{0.36} \\
	  & Binning & 20.00 & 36.0 \\
    \midrule
	  Bark texture & Cake wavelet & 0.70 & 0.79 \\
	  & Ridge & \textbf{0.34} & \textbf{0.36} \\
	  & Binning & 20.00 & 18.00 \\
    \bottomrule
    \end{tabular}
\end{table}

\section{Conclusion}

In this work, we introduce a new numerical method for anisotropic image analysis.
We show experimentally that it is able to retrieve the angular profile of synthetic textures, and that it is robust to numerical rotations on both synthetic textures and real-world images.
Our comparisons also show that the design choices in our method are necessary for robust numerical anisotropic image analysis.
We believe that thanks to its flexibility, our method could be applied in broad image processing settings, including in deep neural networks.

\printbibliography

\begin{table}[t!]
    \centering
    \caption{\textbf{Robustness to noise on angular registration.}
    Each time, both images are corrupted with a Gaussian noise with a SNR of 25 dB.
    The metrics are in degrees, and the best is in \textbf{bold}.
    }
    \label{tab:robustness}
    \setlength{\tabcolsep}{7pt} 
    \begin{tabular}{llcccccc}
    \toprule
    Image & Method & Reg. err. $\downarrow$ & Equiv. err. $\downarrow$ \\
	\midrule
    CT scan & Cake wavelet & \textbf{0.02} & 0.47 \\
    & Ridge & 0.02 & \textbf{0.43} \\
    & Binning & 20.00 & 36.00 \\
    \midrule
    Bark texture & Cake wavelet & 0.70 & 0.83 \\
    & Ridge & \textbf{0.34} & \textbf{0.36} \\
    & Binning & 20.00 & 18.00 \\
    \bottomrule
\end{tabular}

\end{table}
\end{document}